\documentclass{bioinfo}
\usepackage{graphicx}
\usepackage{amssymb}
\usepackage{amsmath}
\usepackage{epstopdf}
\usepackage{color}
\usepackage{verbatim}
\usepackage{algorithm}
\usepackage{algorithmic}
\usepackage{subfig}
\usepackage{multirow}
\usepackage{llncsdoc}

\newtheorem{theorem}{Theorem}[section]
\newtheorem{lemma}[theorem]{Lemma}

\newtheorem{problem}[theorem]{Problem}

\def\quad{ ~ }
\def\qquad{ ~~ }

\newcommand\shrink[1]{}

\copyrightyear{2005}
\pubyear{2005}

\begin{document}
\firstpage{1}
\title{IPED2: Inheritance Path based Pedigree Reconstruction Algorithm for Complicated Pedigrees}

\author[D He \textit{et~al}]{Dan He\,$^{1,*}$, Zhanyong Wang\,$^{2}$, Laxmi Parida\,$^{1}$ and Eleazar Eskin $^{2}$}
\address{$^{1}$IBM T.J. Watson Research, Yorktown Heights, NY.\\
$^{2}$Dept. of Computer Science, University of California Los Angeles, Los Angeles, CA 90095, USA.}
\history{Received on XXXXX; revised on XXXXX; accepted on XXXXX}
\editor{Associate Editor: XXXXXXX}
\maketitle

\begin{abstract}

\section{Motivation:}
Reconstruction of family trees, or pedigree reconstruction, for a group of individuals is a fundamental problem in genetics. The problem is known to be NP-hard even for datasets known to only contain siblings. Some recent methods have been developed to accurately and efficiently reconstruct pedigrees. These methods, however, still consider relatively simple pedigrees, for example, they are not able to handle half-sibling situations where a pair of individuals only share one parent. 

\section{Results:}
In this work, we propose an efficient method, IPED2, based on our previous work, which specifically targets  reconstruction of complicated pedigrees that include half-siblings. We note that the presence of half-siblings makes the reconstruction problem significantly more challenging which is why previous methods exclude the possibility of half-siblings. We proposed a novel model as well as an efficient graph algorithm and experiments show that our algorithm achieves relatively accurate reconstruction. To our knowledge, this is the first method that is able to handle pedigree reconstruction based on genotype data only when half-sibling exists in any generation of the pedigree.

\section{Availability:}
\href{http://www.cs.ucla.edu/~danhe/Software/IPED2.html}{http://www.cs.ucla.edu/$\sim$danhe/Software/IPED2.html}
\section{Contact:} \href{dhe@us.ibm.com}{dhe@us.ibm.com}
\end{abstract}

\section{Introduction}

A pedigree, or family tree, is a diagram which represents genetic relationships between individuals of a family. Since a pedigree defines the genetic relationships between individuals, a pedigree provides a model to compute the inheritance probability for the observed genotype data as all possible inheritance options for an allele in an individual are encoded in the pedigree.  Prediction of pedigrees have been involved in a large number of research studies \cite{coop2008high,ng2009exome,behar2007genographic,servin2004toward,he2013iped} . Therefore, pedigree inference plays an important role in population genetics. 


The pedigree reconstruction problem using genotype data is to reconstruct the unknown pedigree of a set of individuals given genotype data or haplotypes for these individuals. The problem is very challenging in that there are exponential number of possible pedigree graphs and the number of unknown ancestors can be very large as the height of the pedigree increases. Indeed even constructing sibling relationships is known to be NP-hard \cite{kirkpatrick2011pedigree}. The pedigree reconstruction problem has a long history \cite{elston1971general,lander1987construction,abecasis2002merlin,fishelson2005maximum,li2010efficient,sobel1996descent}.

In this work, we focused on reconstruction methods using genotype data. There are existing methods to infer parentage and sibship \cite{jones2010colony,sheikh2010combinatorial}, using maximal likelihood estimation or combinatorial approaches. However, they are not targeting reconstructing the whole pedigree. Various methods \cite{thompson1986pedigree,thatte2008reconstructing,kirkpatrick2011pedigree,he2013iped} have been proposed for automatically reconstructing pedigree using genotype data. The most recent progress on the pedigree reconstruction problem is generation-by-generation reconstruction approaches \cite{kirkpatrick2011pedigree,he2013iped}. The pedigree is reconstructed backwards in time, one generation at a time. The input of these approaches is the set of extant individuals with haplotype and identity-by-descent (IBD) information available. IBD information encodes the haplotype segments in two different haplotypes which have been inherited from the same ancestor. At each generation, a compatibility graph is constructed, where the nodes are individuals and the edges indicate the pair of individuals which could be siblings. The edges are defined via a statistical test such that an edge is constructed only when the test score between the pair of individuals is less than a pre-defined threshold. Sibling sets are identified in the compatibility graph using a Max-clique algorithm iteratively to partition the graph into disjoint sets of vertices. The vertices in the same set have edges connecting to all the other vertices of the same set. Kirkpatrick et al. \cite{kirkpatrick2011pedigree} conducted a sampling algorithm for the statistical test. The method is shown to be effective for outbreeding case but inefficient for inbreeding case. He et al. \cite{he2013iped} proposed an inheritance path based algorithm, IPED, which is able to conduct the statistical test efficiently using a dynamic programming approach. IPED is shown to be efficient for both outbreeding and inbreeding cases.

The two approaches \cite{kirkpatrick2011pedigree} and \cite{he2013iped} only considered pedigrees with simple structure. They can not handle cases where the pedigrees have more complicated structure. A common structure is \textit{half-sibling}, where two children share only one parent. The two approaches can not handle half-siblings, which is indeed a common scenario. This scenario is challenging for two reasons. First, half-siblings and siblings are genetically hard to be distinguished. Second, one individual can be involved in both sibling and half-sibling relationships. As we will show later, due to reconstruction errors, these relationships may conflict with each other. And these relationships also define a set of constraints that we need to follow when we create parents for the individuals.


In this work, we proposed a novel method IPED2 to address the above problem. We proposed a new statistical test to detect half-sibling relationships and a new graph-based algorithm to reconstruct the pedigree when half-sibling is allowed. Our experiments show that IPED2 has better performance for cases where there are half-siblings. To our knowledge, this is the first method that can reconstruct pedigrees with half-siblings using just genotype data.

\section{Preliminaries}

\subsection{Pedigree Graphs}

\begin{figure}
   \centering
   \includegraphics[width=0.4 \textwidth]{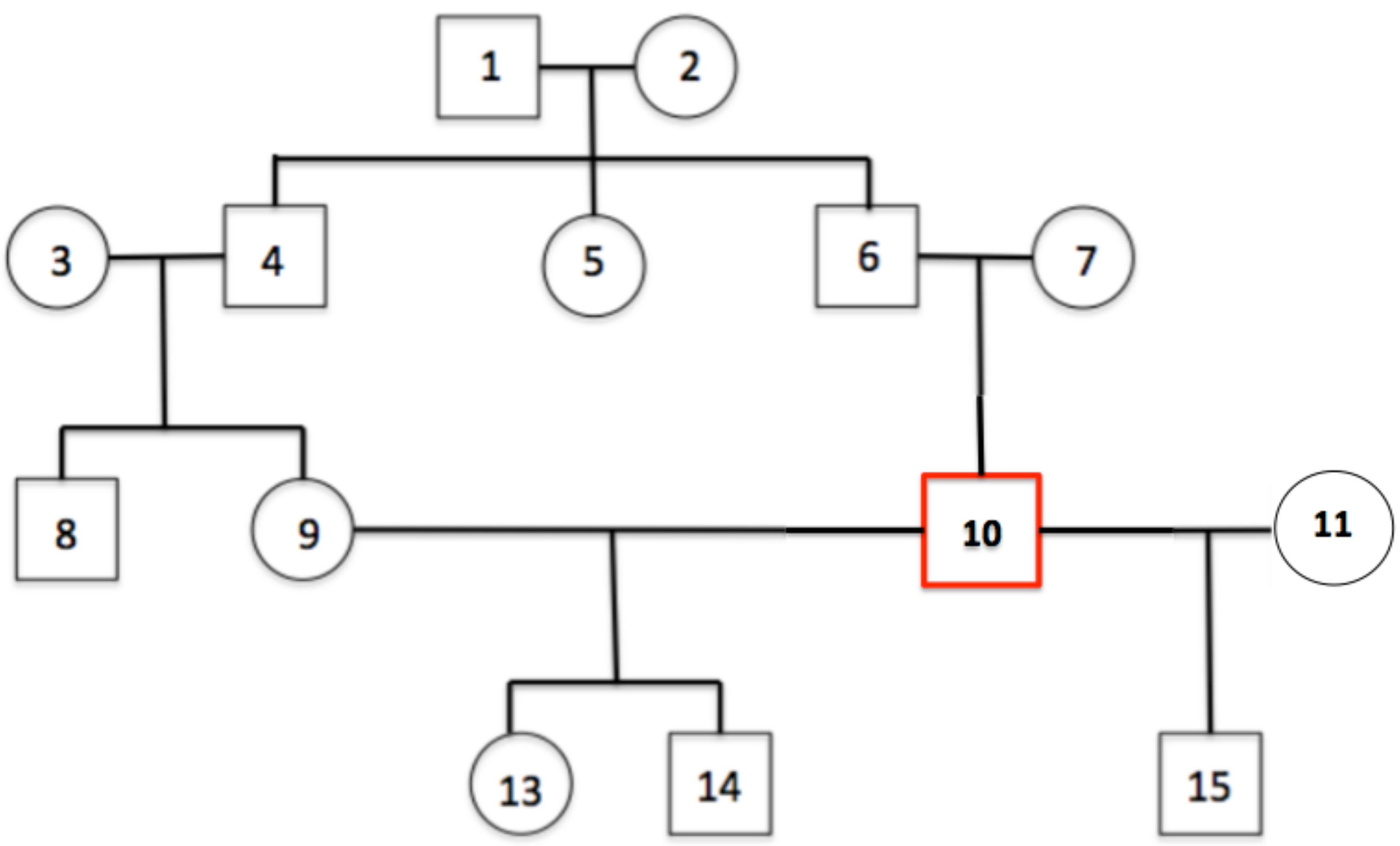}
   \caption{An example of pedigree graph with half-siblings.}
\label{pedigree}
\end{figure}

A pedigree graph consists of nodes and edges where nodes are diploid individuals and edges are between parents and children. Circle nodes are females and boxes are males. An example of pedigree graph is shown in Figure \ref{pedigree}. Parent nodes are also called \textit{founders}.  In the example, individual 13,14,15 are \textit{extant individuals} and their founders are individuals 9, 10 and 10, 11 respectively. \textit{Outbreeding} means an individual mates with another individual from different family. In the example, 3,4 and 6, 7 are both outbreeding cases. \textit{Inbreeding} means an individual mates with another individual from the same family. In the example, 9, 10 is inbreeding case. We can see inbreeding case is usually more complicated as an individual can inherit from his ancestors in multiple ways. For example, 13, 14 can inherit from 1, 2 in two ways but 15 can inherit from 1,2 in only one way.

As we only have extant individuals and we reconstruct ancestors of them, the pedigree is reconstructed backwards in time. We use the same notion of generations in \cite{kirkpatrick2011pedigree}, namely generations are numbered backwards in time, with larger numbers being older generations. Every individual in the graph is associated with a generation $g$. All the extant individuals are associated with $g$=1 and their direct parents are associated with generation $g$=2. The \textit{height} of a pedigree is the biggest $g$. We define \textit{inheritance path}  between a child and his ancestor the same as the one defined in \cite{li2010efficient}, namely as a path between the two corresponding nodes in the pedigree graph. For example, the inheritance path between 1 and 15 consists of nodes 1-6-10-15. There are two inheritance paths between 1 and 13: 1-4-9-13 and 1-6-10-13. Also we assume the inheritance paths are not directed. \textit{Half-sibling} is the case where two children share only one common parent. Node 10 is shared by 13, 14, 15. As 13, 14 also share the other parent 9, they are siblings. The two pairs 13,15 and 14,15 are half-siblings as they only share one common parent. 

The \textit{distance} between parent and children in the pedigree graph is one and the distance between siblings in the pedigree graph is two. \textit{Meiosis} is a type of cell division in which a nucleus divides into four daughter nuclei, each containing half the chromosome number of the parent nucleus. The number of meioses $M$ between a pair of individuals can be computed from the pedigree graph as the distance between them. Therefore for a pair of siblings $M = 2$.  As we will show later, $M$ will be used to evaluate the relationships between individuals so pedigree graph is a critical data structure. 

\subsection{Identical-by-Descent (IBD)}
\begin{figure}
   \centering
   \includegraphics[width=0.15 \textwidth]{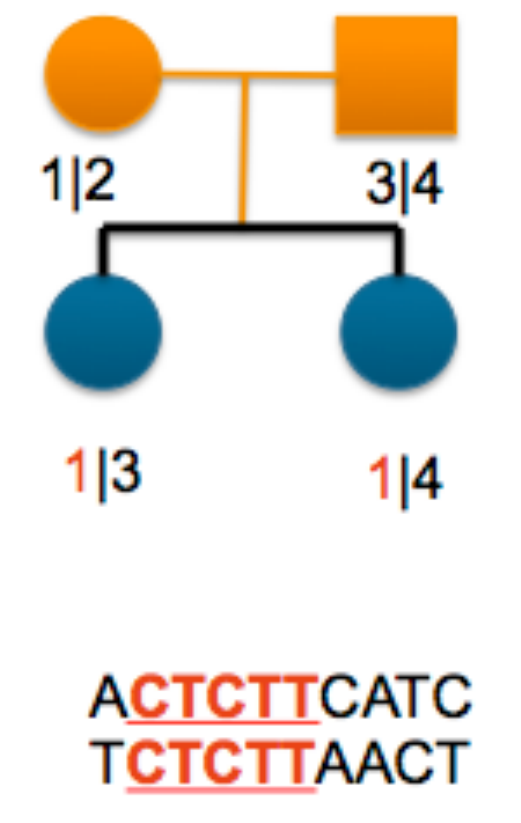}
   \caption{Examples of IBD and IBD tracts.}
\label{IBD}
\end{figure}

Two or more alleles are identical-by-descent (IBD) if they are identical copies of the same ancestral allele. As we can see in Figure \ref{IBD}, the top plot, the two parents have alleles 1,2 and 3,4 respectively and the two children inherit one allele from the father and one allele from the mother. They inherit the same allele 1 from the mother. Therefore the two alleles 1 of the two children are IBD. IBD tracts are consecutive alleles in the genome shared between two or more individuals if they inherit identical nucleotide sequences in the regions from common ancestor. In Figure \ref{IBD}, the bottom plot, we show an example of IBD tracts as the two highlighted sub-sequences.

Many methods have been developed for IBD tracts detection \cite{albrechtsen2009relatedness}~\cite{browning2011fast}~\cite{gusev2009whole}~\cite{purcell2007plink}~\cite{he2013ibd}. Most of these methods are based on haplotype, which is the concatenation of the alleles on the same chromosome. As identical sequences of alleles can happen by chance, the longer the IBD tract is, the more likely it is truly by descent, or truly due to inheritance. As there are two chromosomes in the genome, there are two corresponding haplotypes. Given a pair of haplotypes, we can identify all the IBD tracts between them and measure the average length of the IBD tracts. In this work, we use Beagle \cite{browning2011fast} to identify IBD tracts. 
\\
\\

\subsection{IPED}

Our previous method IPED \cite{he2013iped} is the only known algorithm scalable to large pedigrees with reasonable accuracy for both outbreeding and inbreeding cases using genotype data. IPED starts from the extant individuals and reconstructs the pedigree generation by generation backwards in time. For each generation, IPED predicts the pairwise relationships between the individuals at the current generation and create parents for them according to their relationships. When IPED evaluates the pairwise relationships for a pair of individuals, it considers the pairwise IBD length for their extant descendants, namely the leaf individuals in the pedigree. A statistical test is then conducted on the two individuals to determine if they are siblings or not siblings. The test is different for extant individuals and ancestral individuals. For a pair of extant individuals $i,j$, IPED conducts a statistical test and computes a score $v_{i,j}$ for both sibling case and first-cousin case using Equation \ref{score_extant} and determines $i,j$ are siblings if the score for sibling case is lower. 

{\small
\begin{eqnarray}
v_{i,j} = \frac{\big(estimate(IBD_{i,j}) - E(IBD_{i,j})\big)^2}{var(IBD_{i,j})}
\label{score_extant}
\end{eqnarray}}

\noindent where $estimate(IBD_{i,j})$ is the estimated IBD length between individuals $i$ and $j$, $E(IBD_{i,j})$ is the expected IBD length between $i$ and $j$, $var(IBD_{i,j})$ is the variance of the IBD length between $i$ and $j$. $estimate(IBD_{i,j})$ can be computed easily given genotypes or haplotypes of individual $i$ and $j$. As recombination occurs in meioses,
it is shown \cite{donnelly1983probability} that  the length of IBD between $i$ and $j$ follows an exponential distribution $exp(Mr)$, where $M$ is the number of meioses between $i$ and $j$, $r$ is the recombination rate which is set as $10^{-8}$, namely the probability for recombination occurs at any loci is $10^{-8}$.  Therefore, $E(IBD_{i,j})$ and $var(IBD_{i,j})$ are computed as the following:

{\small \begin{eqnarray}
E(IBD_{i,j}) & = &\frac{1}{M \times r} \\
var(IBD_{i,j}) & = &\frac{1}{(M \times r)^2}
\label{statistic}
\end{eqnarray}}

\noindent The number of meioses $M$ between a pair of individuals can be computed from the pedigree graph. For outbreeding case, $M$ = $2(g-1)$ where $g$ is the generation. So for extant individuals, as we are reconstructing the second generation, $g = 2$ and thus $M = 2$.

To determine the relationship of a pair of ancestral individuals, IPED uses a similar strategy as the one in \cite{kirkpatrick2011pedigree}. Assuming individuals $k$ and $l$ are at generation $g > 1$. The sets of all extant descendants of $k$ and $l$ are $K$ and $L$, respectively. IPED computes a score $v_{k,l}$ between $k$ and $l$ as

{\small \begin{eqnarray}
v_{k,l} &= &\frac{1}{|K||L|} \sum_{i \in K} \sum_{j \in L} v_{i,j}\nonumber\\
 &= &\frac{1}{|K||L|} \sum_{i \in K} \sum_{j \in L} \frac{\big(estimate(IBD_{i,j}) - E(IBD_{i,j})\big)^2}{var(IBD_{i,j})}
\label{score}
\end{eqnarray}}

\noindent where $|K|$ is the size of $K$, the number of extant descendants of $k$, $i \in K$ is an extant individual in $K$, $v_{i,j}$ is computed via Equation \ref{score_extant}.  Again, IPED computes $v_{k,l}$ for both sibling case and first-cousin case and determine $k,l$ are siblings if the score for sibling case is lower.

One of the challenges in this approach is to compute the expected IBD length between a pair of extant individuals efficiently, in the presence of inbreeding. IPED considers the \textit{inheritance paths} between the ancestor and the extant individuals, where each inheritance path corresponds to one path in the pedigree between the ancestor and the extant individual. The benefit of inheritance paths is the distance of a pair of extant individuals can be computed as the sum of the inheritance paths between the two individuals and their common ancestor. If we know all the inheritance paths from the ancestor to the extant individuals, we can compute the probability that an allele of the extant individual is inherited from the ancestor by using the average length of the inheritance paths. The probability can be further utilized to compute the expected average IBD length between a pair of extant individuals 


However, the number of inheritance paths can be exponential. Although the number of inheritance paths can be exponential, their lengths are bounded by the height of the pedigree. Therefore IPED uses a hash data structure \textit{IPP} (Inheritance Path Pair) to hash all the inheritance paths of the same length into a bucket and the number of buckets is bounded by the height of the pedigree and thus is usually small. IPPs are in the form of (length of inheritance path, number of inheritance paths with such length) and they are saved for each individual. A dynamic programming algorithm is developed to populate the hash table of the individuals generation by generation. By doing this, IPED avoids redundant computation of the inheritance paths where the entire pedigree needs to be explored repeatedly and thus the dynamic programming algorithm is very efficient.

\section{Methods}

\subsection{IPED2}

IPED is shown to be efficient for large pedigrees and for both outbreeding and inbreeding cases. However, like all existing methods, it considers only relatively simple pedigrees. In reality, the pedigrees can be much more complicated. In this work, we consider a complicated scenario: half-sibling, where two individuals share only one parent, which is indeed a common scenario.

To address the half-sibling scenario, we proposed IPED2, which is also an inheritance path based pedigree reconstruction method. However, it significantly differs from IPED in both the statistical model and the parent reconstruction algorithm as both of them become much more complicated when half-siblings are considered.

\subsubsection{Workflow}

We first show the workflow of IPED2: starting from the extant individuals IPED2 reconstructs the pedigree generation by generation backwards in time. For each generation, IPED2 identifies the sibling and half-sibling relationships between the individuals at the current generation, using a novel statistic test model, and create parents for them according to their relationships using a novel graph algorithm. The algorithm stops till a certain height of generation is reached. Next we show more details of the method.

\subsubsection{Statistical Test}

In IPED, the statistic test compares only two cases: sibling and non-sibling. For non-sibling, at most two individuals are first-cousin so first-cousin is used. In order to handle half-siblings, in IPED2 we need to compare the statistic test results for fours cases: siblings, half-siblings, first-cousins, first-half-cousins and select the case with the smallest score. As in the genome one allele can be inherited from either father or mother and there are two alleles for each SNP, we need to consider inheritance paths for both alleles, where each allele is on one of the two haplotypes. Therefore we need to consider the average IBD length of both haplotypes. For simplicity, we just call the distance between two alleles from two haplotypes as the distance between the two haplotypes. 

Let's first consider extant individuals. For sibling case, the distance between two individuals is 2 for both haplotypes as alleles from both haplotypes are inherited from the same pair of parents and the distance between parent and child is one. For half-sibling case, the distance between two individuals for one haplotype is 2, for the other haplotype is at least 4, as one pair of alleles from the two individuals are inherited from the same parent, the other pair of alleles from the two individuals are inherited at least from the same grandparent. Similarly, for first-cousin case, the distance between two individuals are 4 for both haplotypes. For first-half-cousin case, the distance between two individuals for one haplotype is 4, for the other haplotype is at least 6. 

Assuming individuals $i, j$ both have a pair of haplotypes noted as $(i_1, i_2), (j_1, j_2)$, there are two possible ways to compare them for IBD, namely $[(i_1,j_1), (i_2,j_2)]$ or $[(i_1,j_2), (i_2,j_1)]$. We select the way that maximizes the sum of the averaged IBD length for both haplotypes as the larger the IBD length is, the more likely the reported IBD tracts are true IBD tracts. We apply the following equation:

{\small \begin{eqnarray}
v_{i,j} = \frac{max(v_{i_1,j_1} + v_{i_2,j_2}, v_{i_1,j_2} + v_{i_2,j_1})}{2}
\label{final_score}
\end{eqnarray}}
\noindent where $v_{i_1,j_1}$ is computed according to Equation \ref{score_extant} by considering the estimated IBD between $i_1, j_1$.

For individuals on higher generations, we compute $v_{k,l}$ for every pair of ancestral individuals $k,l$ using Equation \ref{score}. To compute $v_{i,j}$ for the pair of extant individuals $i,j$, we concatenate the inheritance paths from $k$ to $i$ and from $l$ to $j$ and increase the total length of the concatenated path by $t$ according to a test option for the four different cases. For each case the two combinations of haplotype pairs from $i,j$ may be increased by different length. Therefore, without losing generality, assuming $max(v_{i_1,j_1} + v_{i_2,j_2}, v_{i_1,j_2} + v_{i_2,j_1}) = v_{i_1,j_1} + v_{i_2,j_2}$, we have $t_1$ and $t_2$ as the distances between the alleles from the haplotype pairs $i_1,j_1$ and $i_2,j_2$ between $i,j$, respectively. Assuming $estimate(IBD_{i_1,j_1}) \geq estimate(IBD_{i_2,j_2})$ and $t_1 \leq t_2$, then $t_1$ is applied to $v_{i_1,j_1}$ and $t_2$ is applied to $v_{i_2,j_2}$, namely the smaller one of $t_1$ and $t_2$ is applied to the haplotype with larger average IBD length. For sibling case, $t_1$ = $t_2$ = 2. For half-sibling case, $t_1$ = 2, $t_2$ = 4. For first-cousin case, $t_1$ = $t_2$ = 4.  For first-half-cousin case, $t_1$ = 4, $t_2$ = 6. 


To compute the score $v_{i,j}$, we need to compute $E(IBD_{i,j})$ and $V(IBD_{i,j})$ using Formula \ref{statistic}, where the key factor $M$ is the average distance between $i,j$. We conduct a dynamic programming algorithm similar to that of IPED to compute the inheritance paths which is shown to be very efficient for large pedigrees and for both outbreeding and inbreeding scenarios. 

\begin{figure}
   \centering
   \includegraphics[width=0.4\textwidth]{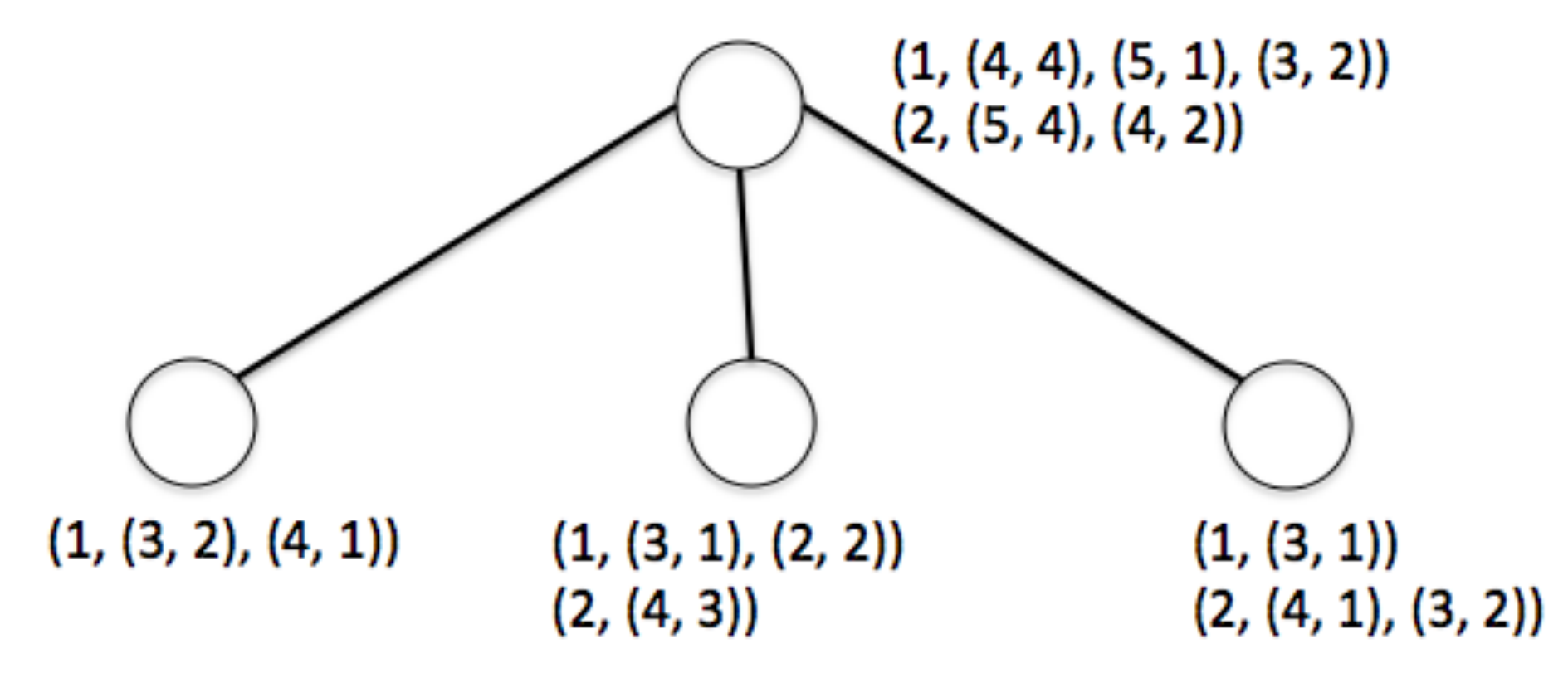}
   \caption{An example of the dynamic programming algorithm. (1, (3, 2), (4, 1)) indicates the corresponding node has a total of 2+1 = 3 inheritance paths to node 1, where 2 of them are of length 3, 1 of them is of length 4. }
\label{DP_example}
\end{figure}


The dynamic programming algorithm starts the reconstruction from generation 2 as generation 1 consists of all the known extant individuals. Then at generation 2, assuming we have a founder $G_2^i$ (without losing generality, assuming he is father) and his $k$ children in generation 1 as $G_1^{i_1}, G_1^{i_2}, \dots, G_1^{i_k}$. Then for every paternal allele of each child, obviously we have 1 possible length 1 inheritance path from the founder. Therefore, we save [$G_1^{i_j}$, (1, 1)] for $G_2^i$ for $1 \leq j \leq k$ where (1, 1) is one IPP of the form (length of inheritance path, number of inheritance paths with such length). Now let's assume we are at generation T, and we are reconstructing generation $T+1$. Again, assuming we have a founder $G_{T+1}^i$ as father and his $k$ children in generation $T$ as $G_T^{i_1}, G_T^{i_2}, \dots, G_T^{i_k}$. We then obtain the IPPs for $G_{T+1}^i$ by merging the IPPs for $G_T^{i_1}, G_T^{i_2}, \dots, G_T^{i_k}$. The recursion is shown as below:
\begin{eqnarray*}
IPP(G_{T+1}^i) =  \sum_{j = 1}^k IPP(G_{T}^{i_j}) + \textbf{1}
\end{eqnarray*}
\noindent where $IPP(G_{T+1}^i)$ is the set of IPPs for node $G_{T+1}^i$. Assuming for $G_T^{i_j}$, we have IPPs \\ $[G_1^t, ((L_{j_1}, N_{j_1}), \dots, (L_{j_m}, N_{j_m}))]$, $IPP(G_{T}^{i_j}) + \textbf{1}$ is to update these pairs as \\ $[G_1^t, (L_{j_1}+1, N_{j_1}), \dots, (L_{j_m}+1, N_{j_m})]$. $IPP(G_{T}^a) + IPP(G_{T}^b)$ is to merge two sets of IPPs. When we merge two pairs $(L_a, N_a)$ and $(L_b, N_b)$, if $L_a = L_b$, we obtain a merged pair $(L_a, N_a + N_b)$. Otherwise we keep the two pairs. Therefore, after the merge, we obtain [$G_1^t, ((L_1, N_1), \dots, (L_m, N_m))]$ for each extant individual $G_1^t$ who is the descendant of $G_{T+1}^i$, where $L_1, \dots, L_m$ are all unique and $m \leq T+1$.
The summation ($\sum_{}^{}$) is similarly defined as the repeated merging operation over multiple sets of IPPs.




An example of the dynamic programming algorithm is shown in Figure \ref{DP_example}.  As we can see in the example, when we merge the IPPs, we increase the length of the paths by 1 and add the number for the paths of the same length. The complexity of this dynamic programming algorithm is $O(E \times k \times H)$ where $E$ is the number of extant individuals, $k$ is the number of direct children for each founder, $H$ is the height of the pedigree. Therefore it is linear time with respect to the height of the pedigree.



Once the inheritance paths are computed, we can compute $M$ using Algorithm \ref{algorithm_ipp}. The average number of meioses $M$ is computed as the average length of the paths between $i, j$. We have $M_1 = ComputeDis(t_1, IPP_i, IPP_j)$ and $M_2 = \\ ComputeDis(t_2, IPP_i, IPP_j)$ where the function \\ $ComputeDis()$ is defined in Algorithm \ref{algorithm_ipp} and $IPP_i$ is the inheritance path pair of $i$. Then we use $M_1$ and $M_2$ to compute $v_{i_1,j_1}$ and $v_{i_2,j_2}$ respectively using Equation \ref{statistic}, which are further used to compute $v_{i,j}$ using Equation \ref{final_score}.

\subsubsection{Construct Parents}
Another challenging problem when half-sibling is considered is to construct parents. To construct parents, IPED builds a sibling graph on each generation where the nodes are individuals and the edges indicate sibling relationships. A maximum clique algorithm is applied to select the cliques of siblings in a greedy manner. In IPED2, we need to consider both sibling and half-sibling relationships and one node can be in both relationships.

\begin{figure}
   \centering
   \includegraphics[width=0.35 \textwidth]{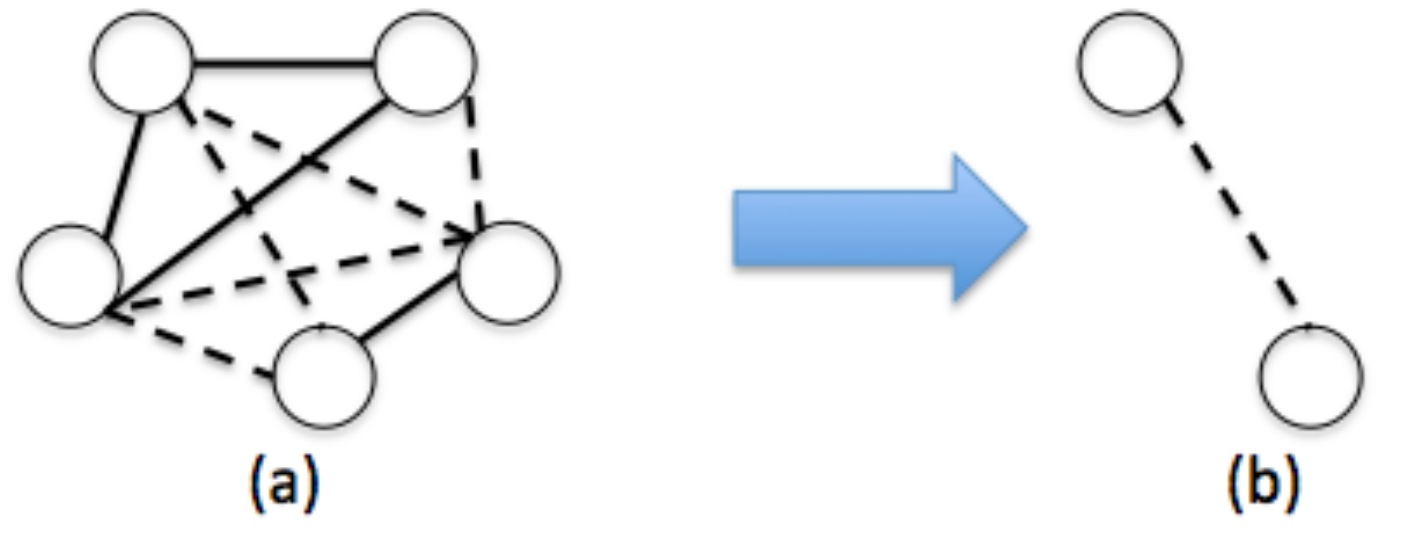}

   \caption{Examples of a relationship graph (a) and the corresponding virtual half-sibling graph (b).}
\label{relationship}
\end{figure}

To address this problem, we build a \textit{relationship} graph, where the nodes are individuals, the edges indicate sibling or half-sibling relationships. As shown in Figure \ref{relationship} (a), we use concrete edges to indicate sibling relationships (we call them \textit{sibling edges}), and dashed edges to indicate half-sibling relationships (we call them \textit{half-sibling edges}). The problem of creating parents for each individual given the graph is equivalent to the following labeling problem:

\begin{problem}
Given a relationship graph, we would like to assign each node with two labels, such that the labels satisfy the \textit{labeling constraints}: (1) Each node has two different labels. (2) A pair of nodes connected by a sibling edge have two identical labels. (3) A pair of nodes connected by a half-sibling edge have one identical label and one different label. (4) A pair of nodes that are not connected by any edge have two different labels.
\label{constraints}
\end{problem}

\begin{algorithm}
\caption{ComputeDis($t, IPP_i, IPP_j$): Once all the inheritance paths are computed, we can calculate the average number of meioses between a pair of nodes $i,j$ given their IPPs \cite{he2013iped}}
\begin{algorithmic}
\REQUIRE $t$ (test options), $IPP_i = [i, ((l_{g_1}, n_{g_1}), \dots, (l_{g_h}, n_{g_h})]$ and $IPP_j = [j, ((l_{k_1}, n_{k_1}), \dots, (l_{k_f}, n_{k_f}))]$
\ENSURE The approximate average path length between $i,j$
\STATE $Length \leftarrow 0$
\STATE $Num \leftarrow 0$
\FOR{$a = 1$ to h}
\FOR{$b = 1$ to f}
\STATE $Num \leftarrow Num + n_{g_a} \times n_{k_b}$
\STATE $Length \leftarrow Length + (l_{g_a} + l_{k_b} + t) \times (n_{g_a} \times n_{k_b})$
\ENDFOR
\ENDFOR
\STATE $approximate~average~path~length \leftarrow \frac{Length}{Num}$
\end{algorithmic}
\label{algorithm_ipp}
\end{algorithm}

\begin{figure}
   \centering
   \includegraphics[width=0.4 \textwidth]{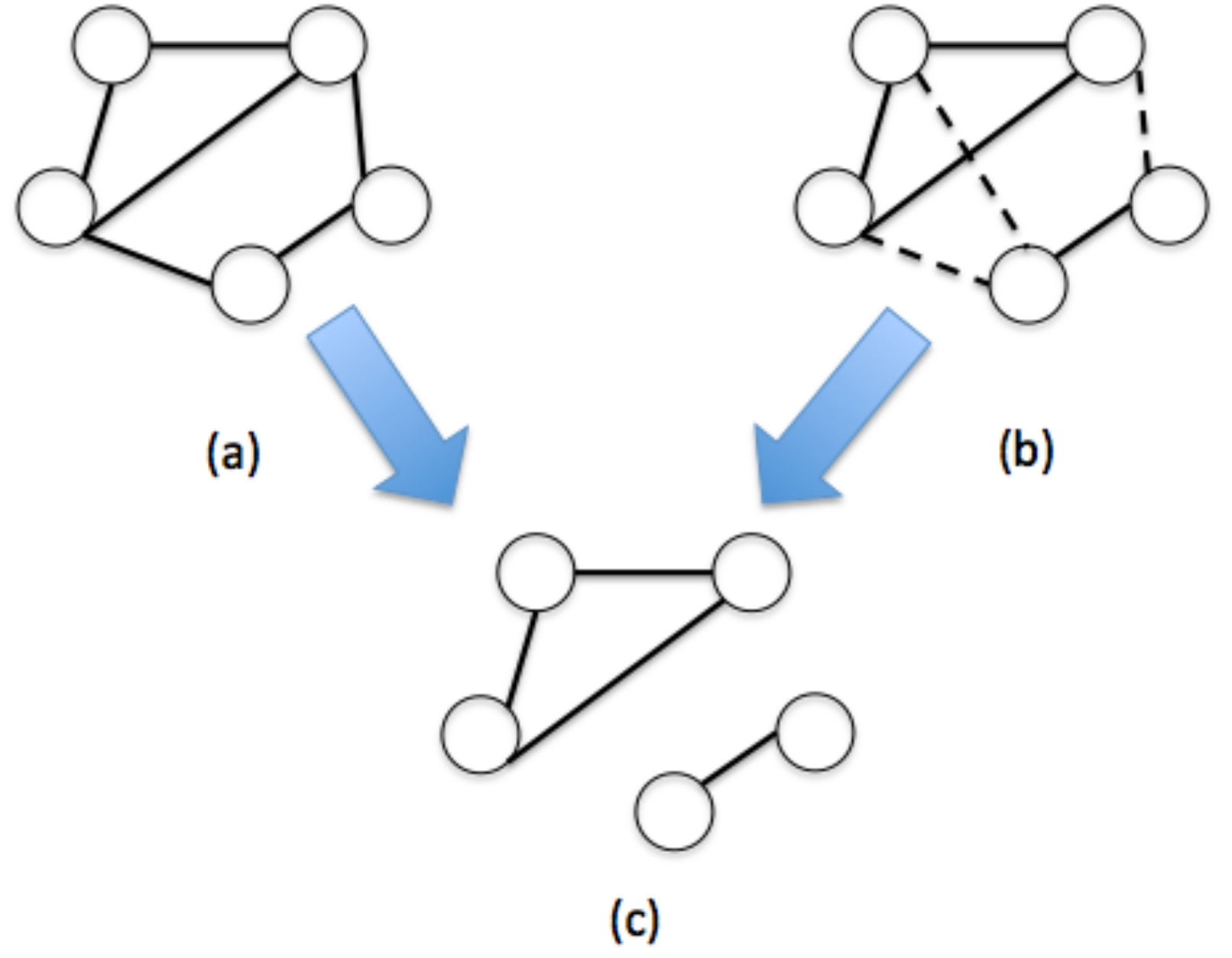}
   \caption{Examples of possible conflicts in relationship graph ((a) and (b)) and their resolution (c).}

\label{conflict}
\end{figure}

From now on we will use ``labels" and ``parents" interchangeably. We call a solution that satisfies the above constrains a \textit{valid} solution. The challenge of the above problem is that a valid solution may not always exist for a given relationship graph. For example, in Figure \ref{conflict} (a), all of the node should have identical labels as they are all connected by sibling edges. However, some of them are not connected directly by a sibling-edge, indicating they should have different labels. Thus there is no valid solution. This is mainly due to the statistical test errors where the tests either report false positive or false negative sibling relationships.

Therefore, in order to guarantee a valid solution, we may need to delete certain edges in the relationship graph to resolve conflict. Notice we delete edges instead of inserting edges to resolve conflicts because it would be complicated to determine which type of edges to insert and we try to be conservative on determining the relationships. As there are exponential number of ways to delete the edges with respect to the number of edges, we propose a revised labeling problem:
\begin{problem}
Given a relationship graph, find a valid labeling solution which requires the minimum number of edge deletions and satisfies all the labeling constraints on the graph after the edge deletions.
\end{problem}

Notice the above problem is an optimization problem and a valid solution always exists. A naive solution is to delete all the edges such that all the nodes are disconnected. Then we can create a pair of unique parents for each node. This solution is valid but obviously not the optimal one.

To solve the above problem, a naive algorithm is to enumerate all possible combinations of the edges to be deleted. Then we can check for each combination if there is a valid labeling solution. This algorithm is optimal but it is of complexity $O(e!)$ where $e$ is the number of edges in the graph and is obviously not feasible for even small graphs.

We observe that in the relationship graph, many edges are irrelevant to the conflicts and deleting them would not help to resolve the conflicts. We also observe that there are two types of conflicts as shown below:

\begin{itemize}
\item Conflicts within connected components by sibling edges. As shown in Figure \ref{conflict} (a), we can see the connected component is not a clique as there are missing edges. As sibling relationships should be transitive, there is a conflict.
\item Conflicts between sibling cliques. As shown in Figure \ref{conflict} (b), if two sibling cliques are connected by at least one half-sibling edge, all pairs of nodes between the two cliques should be connected by a half-sibling edge. Otherwise there is a conflict.
\end{itemize}

\begin{figure}
   \centering
   \includegraphics[width=0.36 \textwidth]{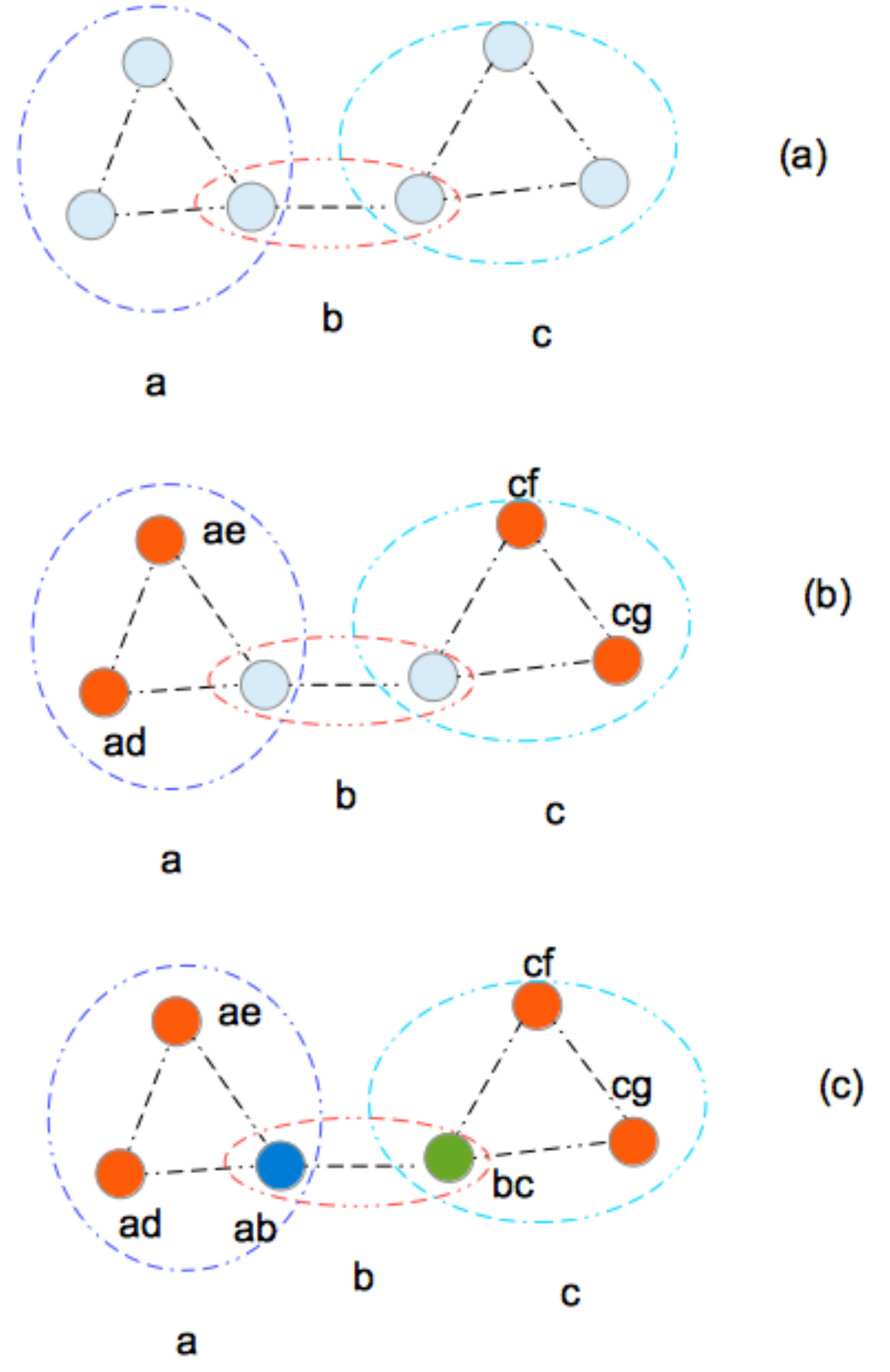}
   \caption{Examples of sequentially labeling the half-sibling graph. Each step of the labeling process is shown.}
\label{reconstruction}
\end{figure}

\begin{table*}[ht]
\caption{The values of the five sets of parameters. Considering the fact that we only have few families in each generation, the true half-sibling rate is much lower than the pre-set parameter. Please find the detailed explanation of the variables in the paper.}
\centering
\begin{tabular}{|c|c|c|c|c|}
\hline
	Parameter Set & Ave. num. of children& Num. of indi. each generation & Rate of half-sibling & Height \\\hline
	1 & 3 & 20 & 0 & 5\\\hline
	2 & 2 & 20 & 0.8 & 5 \\\hline
	3 & 3 & 40 & 0.5 & 5 \\\hline
	4 & 3 & 20 & 0.8 & 10 \\\hline
	5 & 3 & 40 & 0.8 & 10 \\\hline
\end{tabular}
\label{parameters}
\end{table*}

\begin{table*}
\caption{Reconstruction accuracy for IPED2, IPED and COP for parameter sets one, two and three as shown in Table \ref{parameters}. We vary the height of the pedigree.}
\label{first_three}
{\scriptsize \begin{tabular}{|c||c|c|c|c||c|c|c|c||c|c|c|c|}
\hline
    Parameter & \multicolumn{4}{ |c| }{One} & \multicolumn{4}{ |c| }{Two} & \multicolumn{4}{ |c| }{Three} \\\hline
	Height & Family Size & IPED2 & IPED & COP & Family Size & IPED2 & IPED & COP & Family Size & IPED2 & IPED & COP\\\hline
	g = 2 & 32 & 0.962 & 0.995 & 0.83 & 34 & 0.976 & 0.964 & 0.842 & 63 & 0.972 & 0.968 & 0.892\\\hline
	g = 3 & 42 & 0.780 & 0.789 & 0.428 & 50 & 0.684 & 0.631 & 0.422 & 82 & 0.750 & 0.721 & 0.576 \\\hline
	g = 4 & 52 & 0.714 & 0.772 &  0.323 & 62 & 0.618 & 0.569 &  0.339 & 102 & 0.624 & 0.609 &  0.321 \\\hline
	g = 5 & 64 & 0.704 & 0.771 &  0.322 & 72 & 0.603 & 0.569 &  0.337 & 122 & 0.616 & 0.601 &  0.310\\\hline
\end{tabular}}
\end{table*}

\begin{table*}
\parbox{1.0\linewidth}{
\caption{Reconstruction accuracy for IPED2, IPED and COP for parameter sets four and five as shown in Table \ref{parameters}.}
\centering
\begin{tabular}{|c|c|c|c|c|c|}
 \hline
	Parameter set & Family Size & Generation & IPED2 & IPED & COP \\\hline
	4 & 101 & 10 & 0.333 & 0.284 & 0.275\\\hline
	5 & 232 & 10 & 0.240 & 0.214 & 0.163\\\hline
\end{tabular}
\label{deep_pedigree}
}
\end{table*}

Notice the cliques can be of size one, namely it has only one node. Therefore, we can first identify the cliques involved in the conflicts and remove only those related edges to resolve the conflicts. For conflicts among sibling cliques, we conduct a greedy strategy where we select the maximal clique and then remove all the nodes and edges of the clique. We then identify the next maximal clique in the remaining graph and we repeat the process until there is no node left. An example is shown in Figure \ref{conflict} (a). For conflicts between sibling cliques, we check if all pairs of nodes between the two cliques are connected by a half-sibling edge. If not, we remove all the half-sibling edges between the two cliques. An example is shown in Figure \ref{conflict} (b). So the two examples in Figure \ref{conflict} are both reduced to Figure \ref{conflict} (c) once we resolve the conflicts.

Once we resolve all the conflicts, there is always a valid labeling solution for the relationship graph. The problem is how should we label the nodes. For nodes that are involved in the sibling cliques, it's relatively easy: we just need to create a pair of unique labels for each clique and all the nodes share the same pair of labels. However, if one sibling clique is connected to another sibling clique by half-sibling edges, the two sibling cliques should share one common parent. Therefore, when we create a pair of parents, we not only need to check the sibling relationships, but also we need to check the half-sibling relationships. As an individual can have many different relationships, the problems becomes very complicated.

In order to develop an efficient labeling algorithm, we build a ``virtual half-sibling'' graph from the relationship graph based on the motivation that all the nodes in the same sibling clique should share the same pair of labels. Thus we merge every sibling clique into a ``virtual" node and connect these ``virtual" nodes by ``virtual" half-sibling edges if the two sibling cliques are still connected by half-sibling edges after we have resolved all the conflicts. An example is shown in Figure \ref{relationship} (b). Now the problem is reduced to labeling a graph where there are only half-sibling edges and thus becomes easier.

We first show the following three lemmas.

\begin{lemma}
Half-sibling relationships are not transitive.
\label{non-transitive}
\end{lemma}
\noindent \textbf{Proof:} As shown in Figure \ref{pedigree}, nodes 13 and 15 are half-siblings and nodes 14 and 15 are half-siblings. However, nodes 13 and 14 are siblings instead of half-siblings. Therefore, half-sibling relationships are not transitive. We can not simply identify maximal cliques in the half-sibling graph when we create parents for the nodes.

\begin{lemma}
In the half-sibling graph, all the nodes in the same clique of size greater than one share one and only one common parent.
\label{half-sibling-constraint}
\end{lemma}

\noindent \textbf{Proof:} For a clique of size two, as the two nodes are half-siblings, they must share the same one common parent and only this parent. For a clique of size larger than two, we start with clique of size three. For example, we have three individuals $A,B,C$ who are in the same clique. Assuming $A$ has parent 1,2 and $B$ has parent 2,3. As $C$ is half-sibling of both $A$ and $B$, $C$ needs to share one parent with $A$ and one parent with $B$. The only possible parents for $C$ are (1,3) or (2,4) where 4 is a unique parent for $C$. However, as 2 mates with both 1 and 3, 1 and 3 need to be of the same sex and they can not mate. So the pair of parents (1,3) would not happen in reality. Thus the only valid solution is (2,4) and $A,B,C$ share one and only one common parent. For a clique of size larger than three, as it can be decomposed into sub-cliques of size three, all the nodes need to share the same one common parent and only the parent.

\begin{lemma}
One individual can be only in at most two half-sibling cliques of size greater than one.
\label{two-cliques}
\end{lemma}
\noindent \textbf{Proof:} For an individual to involve in a half-sibling clique, one of the parents need to mate with multiple individuals. As one individual has only two parents, at most both parents mate with multiple individuals. Therefore the individual can only involve in at most two half-sibling cliques.

Based on the above lemmas, we introduce a labeling algorithm which is guaranteed to be valid. We first apply the well known Bron-Kerbosch algorithm \cite{bron1973algorithm} on the ``virtual half-sibling" graph, which identifies all the maximum cliques of the graph. The time complexity of the algorithm is $O(N^3)$ where $N$ is the number of nodes in the graph. Then for each maximum clique, we assign an unique common label, which will be the shared parents of the clique. Next we take each individual of each clique, if the individual is in only one clique, we assign a unique label to the individual as the other parent of this individual. If the individual is in two cliques, his parents are the combination of the two unique labels of the two cliques. According to Lemma \ref{two-cliques}, an individual can be in at most two cliques. Thus the above labeling algorithm is guaranteed to generate valid labels. 

Notice the above algorithm has complexity $O(N^3 + N)$ where $O(N^3)$ is the complexity to identify all maximum cliques and $O(N)$ is the complexity to assign labels for each individual. This complexity is high if the number of nodes $N$ in the graph is large. In reality, the number of individuals on each generation is relatively small, and $N$ is the number of ``virtual half-sibling" cliques. Thus $N$ is a relatively small number. So the algorithm is applicable to relatively large pedigrees. 

For illustration purpose, we show an example in Figure \ref{reconstruction}. Clearly the graph contains 3 cliques. Thus in Figure \ref{reconstruction} (a), we assign a unique common label to each clique as $a, b, c$, respectively. Then in Figure \ref{reconstruction} (b), in each clique, we identify the nodes that belong to only this clique. These four nodes are marked with orange color. We assign one unique label to each such node. Combined with their clique labels, we obtain the pair of labels for these four nodes as $ad, ae, cf, cg$. In Figure \ref{reconstruction} (c), for the blue node in both the clique with common label $a$ and common label $b$, we pair the two labels to obtain $ab$. For the green node in both the clique with common label $b$ and common label $c$, we pair the two labels to obtain $bc$. Thus we obtain a set of valid labels for this graph.

\subsection{Performance Evaluation}

To evaluate the performance of IPED2, we use the same metric used in \cite{he2013iped}:

{\scriptsize \begin{eqnarray*}
accuracy(R, O) & = &\frac{\sum_{i \in E, j \in E} F(R_{i,j}, O_{i,j})}{|E|^2} \\
  F(R_{i,j}, O_{i,j}) & = & \left\{
  \begin{array}{l l}
    1 & \quad \text{if $R_{i,j} = O_{i,j}$}\\
    0 & \quad \text{otherwise}\\
  \end{array} \right.
\end{eqnarray*}}
\noindent where $R$ is the reconstructed pedigree, $O$ is the original pedigree, $E$ is the set of extant individuals, $|E|$ is the number of extant individuals, $R_{i,j}$ is the distance of individual $i$ and $j$ in pedigree $R$ and $R_{i,j} = \infty$ if $i,j$ are not connected in the pedigree graph. Notice if there are multiple paths between $i$ and $j$ in $R$, we select the shortest path.


\section{Results}


In order to test the performance of our method on complicated pedigrees, we use simulated pedigrees with different parameter settings and instead of genotype data, we simulate haplotypes directly. The haplotypes of the individuals are generated according to the Wright-Fisher Model \cite{press2011wright} with monogamy. The model takes parameters for a fixed populations size, a Poisson number of offspring and a number of generations (or the height of pedigree). We consider identical regions of length greater than 1Mb as IBD regions. We compare IPED2 with IPED \cite{he2013iped} and COP \cite{kirkpatrick2011pedigree}, the only two known algorithms that are scalable to relatively large pedigrees. COP is designed for only outbred pedigrees because the related algorithm which can handle inbreeding, CIP, is not scalable to large pedigrees. All the experiments are done on a 2.4GHz Intel Dual Core machine with 4G memory.


We tested 5 sets of parameters, summarized in Table \ref{parameters}, by varying every parameter of the Wright-Fisher Model. The parameters include the average number of children of each family, the approximate number of individuals of each generation, the rate of half-sibling, the height, or the number of generations of the pedigree. Note that the parameter that represents the number of individuals per generation is only a parameter in the Wright Fisher model and the actual number of individuals generated in the simulation may differ. The rate of half-sibling is the probability to have ``wife-husband-wife"(or ``husband-wife-husband") triples in each generation.  According to the Wright-Fisher model, the number of individuals in each generation is a constant number. So, in order to simulate the next generation with the same number of individuals, we generate children from the current generation by forming couples(or triples) incrementally until we have enough children. If the last family is a triple, it is not rare that we obtain enough children from only one couple and the half-siblings from this family is not selected into the next generation. Considering the fact that we only have few families in each generation, the true half-sibling rate is much lower than the pre-set parameter. For each set of parameters, we randomly generate 10 pedigrees and report results averaged over the 10 pedigrees.

In Table \ref{first_three}, we show reconstruction accuracy for IPED2, IPED and COP for parameter sets $1$, $2$ and $3$. For parameter set 1, the rate of half-sibling is 0, namely there is no half-sibling in the pedigrees. We can see that the accuracy of IPED2 and IPED are both much better than that of COP. IPED2 has slightly worse accuracy than that of IPED, due to the false positive half-sibling relationships. However, when the rate of half-sibling is not 0, for parameter sets 2 and 3, IPED2 has better accuracy than that of IPED, indicating that IPED2 is able to improve the reconstruction accuracy by effectively capturing half-sibling relationships.

We also tested the performance of IPED2, IPED and COP for deep pedigrees with 10 generations, with parameter settings $4$ and $5$. The results are shown in Table \ref{deep_pedigree}. IPED2 again achieves much better accuracy than those of IPED and COP. For both parameter settings, both IPED2 and IPED outperform COP significantly. All three methods finish in seconds even for the deep pedigrees, indicating IPED2 is scalable to large and complicated pedigrees.

Finally we show the number of reconstructed half-sibling relationships and the real half-sibling relationships for parameter set 2 to 5 in Table \ref{half_sib_number}. We can see that IPED2 missed about half of the true half-sibling relationships. This might be because we are relatively conservative at claiming half-sibling relationships by deleting all suspicious half-sibling relationships from the relationship graphs. In our future work, we would like to develop more effective strategies to identify half-sibling relationships.

\begin{table}
\parbox{.9\linewidth}{
\caption{Number of reconstructed half-siblings vs. real half-siblings for the last four different parameter sets.}
\centering
\begin{tabular}{|c|c|c|} \hline
	Parameter Set & Reconstructed & Real \\\hline
	2 & 32 & 80 \\\hline
	3 & 62 & 150  \\\hline
	4 & 42 & 122  \\\hline
	5 & 111 & 220  \\\hline
\end{tabular}
\label{half_sib_number}
}
\end{table}

\section{Discussion and Future Work}
We have presented a new efficient method for pedigree reconstruction, IPED2, which is designed to handle complex pedigrees. IPED2 is particularly designed to handle inbreeding and the presence of half-siblings which to our knowledge, previous methods could not address when genotype data is used. It utilizes a novel statistic test model and a novel graph algorithm to handle half-sibling scenario. We show that our method compares favorably under a wide range of simulation scenarios to previous methods such as IPED and COP and it is scalable to large pedigrees. In our future work, we would like to explore a better way to handle the conflicts in the relationship graph to capture more true half-sibling relationships. For example, we would like to consider more actions such as insertion, deletion and replacement to resolve the conflicts. We would also like to consider cases where genotypes of ancestral individuals are known and genotypes of extant individuals that are not on the lowest generations are known.

\bibliographystyle{plain}

\bibliography{pedigree}

\begin{thebibliography}{10}

\bibitem{abecasis2002merlin}
G.R. Abecasis, S.S. Cherny, W.O. Cookson, L.R. Cardon, et~al.
\newblock Merlin-rapid analysis of dense genetic maps using sparse gene flow
  trees.
\newblock {\em Nature genetics}, 30(1):97--101, 2002.

\bibitem{albrechtsen2009relatedness}
A.~Albrechtsen, T.~Sand~Korneliussen, I.~Moltke, T.~van Overseem~Hansen, F.C.
  Nielsen, and R.~Nielsen.
\newblock Relatedness mapping and tracts of relatedness for genome-wide data in
  the presence of linkage disequilibrium.
\newblock {\em Genetic epidemiology}, 33(3):266--274, 2009.

\bibitem{behar2007genographic}
D.M. Behar, S.~Rosset, J.~Blue-Smith, O.~Balanovsky, S.~Tzur, D.~Comas, R.J.
  Mitchell, L.~Quintana-Murci, C.~Tyler-Smith, and R.S. Wells.
\newblock The genographic project public participation mitochondrial dna
  database.
\newblock {\em PLoS Genetics}, 3(6):e104, 2007.

\bibitem{bron1973algorithm}
Coen Bron and Joep Kerbosch.
\newblock Algorithm 457: finding all cliques of an undirected graph.
\newblock {\em Communications of the ACM}, 16(9):575--577, 1973.

\bibitem{browning2011fast}
B.L. Browning and S.R. Browning.
\newblock A fast, powerful method for detecting identity by descent.
\newblock {\em The American Journal of Human Genetics}, 88(2):173--182, 2011.

\bibitem{coop2008high}
G.~Coop, X.~Wen, C.~Ober, J.K. Pritchard, and M.~Przeworski.
\newblock High-resolution mapping of crossovers reveals extensive variation in
  fine-scale recombination patterns among humans.
\newblock {\em Science}, 319(5868):1395--1398, 2008.

\bibitem{donnelly1983probability}
K.P. Donnelly.
\newblock The probability that related individuals share some section of genome
  identical by descent.
\newblock {\em Theoretical Population Biology}, 23(1):34--63, 1983.

\bibitem{elston1971general}
R.C. Elston and J.~Stewart.
\newblock A general model for the genetic analysis of pedigree data.
\newblock {\em Human heredity}, 21(6):523--542, 1971.

\bibitem{fishelson2005maximum}
M.~Fishelson, N.~Dovgolevsky, D.~Geiger, et~al.
\newblock Maximum likelihood haplotyping for general pedigrees.
\newblock {\em Human Heredity}, 59(1):41--60, 2005.

\bibitem{gusev2009whole}
A.~Gusev, J.K. Lowe, M.~Stoffel, M.J. Daly, D.~Altshuler, J.L. Breslow, J.M.
  Friedman, and I.~Pe'er.
\newblock Whole population, genome-wide mapping of hidden relatedness.
\newblock {\em Genome Research}, 19(2):318--326, 2009.

\bibitem{he2013ibd}
Dan He.
\newblock Ibd-groupon: an efficient method for detecting group-wise
  identity-by-descent regions simultaneously in multiple individuals based on
  pairwise ibd relationships.
\newblock {\em Bioinformatics}, 29(13):i162--i170, 2013.

\bibitem{he2013iped}
Dan He, Zhanyong Wang, Buhm Han, Laxmi Parida, and Eleazar Eskin.
\newblock Iped: Inheritance path based pedigree reconstruction algorithm using
  genotype data.
\newblock In {\em Research in Computational Molecular Biology}, pp. 75--87.
  Springer, 2013.

\bibitem{jones2010colony}
Owen~R Jones and Jinliang Wang.
\newblock Colony: a program for parentage and sibship inference from multilocus
  genotype data.
\newblock {\em Molecular Ecology Resources}, 10(3):551--555, 2010.

\bibitem{kirkpatrick2011pedigree}
B.~Kirkpatrick, S.~Li, R.~Karp, and E.~Halperin.
\newblock Pedigree reconstruction using identity by descent.
\newblock In {\em Research in Computational Molecular Biology}, pp. 136--152.
  Springer, 2011.

\bibitem{lander1987construction}
E.S. Lander and P.~Green.
\newblock Construction of multilocus genetic linkage maps in humans.
\newblock {\em Proceedings of the National Academy of Sciences}, 84(8):2363,
  1987.

\bibitem{li2010efficient}
X.~Li, X.~Yin, and J.~Li.
\newblock Efficient identification of identical-by-descent status in pedigrees
  with many untyped individuals.
\newblock {\em Bioinformatics}, 26(12):i191--i198, 2010.

\bibitem{ng2009exome}
S.B. Ng, K.J. Buckingham, C.~Lee, A.W. Bigham, H.K. Tabor, K.M. Dent, C.D.
  Huff, P.T. Shannon, E.W. Jabs, D.A. Nickerson, et~al.
\newblock Exome sequencing identifies the cause of a mendelian disorder.
\newblock {\em Nature genetics}, 42(1):30--35, 2009.

\bibitem{press2011wright}
W.H. Press.
\newblock Wright-fisher models, approximations, and minimum increments of
  evolution.
\newblock 2011.

\bibitem{purcell2007plink}
S.~Purcell, B.~Neale, K.~Todd-Brown, L.~Thomas, M.A.R. Ferreira, D.~Bender,
  J.~Maller, P.~Sklar, P.I.W. De~Bakker, M.J. Daly, et~al.
\newblock Plink: a tool set for whole-genome association and population-based
  linkage analyses.
\newblock {\em The American Journal of Human Genetics}, 81(3):559--575, 2007.

\bibitem{servin2004toward}
B.~Servin, O.C. Martin, M.~M{\'e}zard, et~al.
\newblock Toward a theory of marker-assisted gene pyramiding.
\newblock {\em Genetics}, 168(1):513--523, 2004.

\bibitem{sheikh2010combinatorial}
Saad~I Sheikh, Tanya~Y Berger-Wolf, Ashfaq~A Khokhar, Isabel~C Caballero,
  Mary~V Ashley, Wanpracha Chaovalitwongse, Chun-An Chou, and Bhaskar DasGupta.
\newblock Combinatorial reconstruction of half-sibling groups from
  microsatellite data.
\newblock {\em Journal of Bioinformatics and Computational Biology},
  8(02):337--356, 2010.

\bibitem{sobel1996descent}
E.~Sobel and K.~Lange.
\newblock Descent graphs in pedigree analysis: applications to haplotyping,
  location scores, and marker-sharing statistics.
\newblock {\em American journal of human genetics}, 58(6):1323, 1996.

\bibitem{thatte2008reconstructing}
B.D. Thatte and M.~Steel.
\newblock Reconstructing pedigrees: A stochastic perspective.
\newblock {\em Journal of theoretical biology}, 251(3):440--449, 2008.

\bibitem{thompson1986pedigree}
E.A. Thompson.
\newblock {\em Pedigree analysis in human genetics}.
\newblock Johns Hopkins University Press Baltimore, MD, 1986.

\end{thebibliography}

\end{document}